# AlGaAs/GeSn p-i-n diode interfaced with ultrathin Al$_2$O$_3$

Yang Liu, Yiran Li, Sudip Acharya, Jie Zhou, *Member, IEEE,* Jiarui Gong, *Member, IEEE*, Alireza Abrand, Yi Lu, Daniel Vincent, Samuel Haessly, Parsian K. Mohseni, Shui-Qing Yu, Senior *Member, IEEE,* and Zhenqiang Ma, *Fellow, IEEE*

*Abstract*— **This study presents the fabrication and characterizations of an Al$_{0.3}$Ga$_{0.7}$As/Ge$_{0.87}$Sn$_{0.13}$/GeSn p-i-n double heterostructure (DHS) diode following the grafting approach for enhanced optoelectronic applications. By integrating ultra-thin Al$_2$O$_3$ as a quantum tunneling layer and enhancing interfacial double-side passivation, we achieved a heterostructure with a substantial 1.186 eV conduction band barrier between AlGaAs and GeSn, along with a low interfacial density of states. The diode demonstrated impressive electrical characteristics with high uniformity, including a mean ideality factor of 1.47 and a mean rectification ratio of 2.95×10$^3$ at ±2 V across 326 devices, indicating high-quality device fabrication. Comprehensive electrical characterizations, including C-V and I-V profiling, affirm the diode's capability to provide robust electrical confinement and efficient carrier injection. These properties make the Al$_{0.3}$Ga$_{0.7}$As/Ge$_{0.87}$Sn$_{0.13}$/GeSn DHS a promising candidate for next-generation electrically pumped GeSn lasers, potentially operable at higher temperatures. Our results provide a viable pathway for further advancements in various GeSn-based devices.**

*Index Terms*— **Aluminum gallium arsenide, germanium tin, double heterostructure, grafting, electrical confinement, interface, ultrathin oxide**

## I. INTRODUCTION

Germanium-tin (GeSn) alloy has emerged as a pivotal material within the group-IV semiconductors, offering substantial potential for monolithic optoelectronic integrated circuits in the recent years [1-3]. Its ability to transition from an indirect to a direct bandgap around 8% tin composition[4], positions GeSn as a viable mid-wave infrared light source compatible with CMOS technology on silicon substrates[5]. Since the achievement of the first optically pumped GeSn laser at 90 K in 2015[6], advancements have significantly enhanced the performance of these lasers. GeSn microdisk lasers utilizing tensile strain can now operate at 300 K with an energy density of 550 kW·cm$^{-2}$ [7], alongside other strain-managed film-based devices[8-10].

The development of electrically pumped GeSn lasers has garnered attention due to their potential on-chip applications[11]. The inaugural electrically injected Fabry-Perot GeSn laser, demonstrated in 2020, operated in a pulsed mode, delivering an output of 2.7 mW per facet at 10 K, with a maximum lasing temperature of 100 K[12]. Advances in SiGeSn/GeSn/Ge p-i-n double heterostructures (DHS) have pushed the operational temperature limit to 140 K, with a lasing threshold of 0.756 kA/cm$^2$ at 77 K[13]. SiGeSn is latticed-matched and forms a Type-I band alignment with GeSn active layer, serving as a carrier injection and optical confinement cap[1, 14, 15]. Key challenges to achieving higher operating temperatures and reducing threshold currents have been identified, including metal contact scattering losses, electron confinement in the cap layer, and free carrier absorption[16]. At the cap/active region interface, Si$_{0.03}$Ge$_{0.89}$Sn$_{0.08}$ provides a 114 meV conduction band barrier to Ge$_{0.89}$Sn$_{0.11}$, facilitating electron confinement, while the GeSn buffer serves as the hole barrier. Such DHS confinement works well at low temperatures, but is ineffective as it approaches room temperature due to the thermally excited carrier leakage[17]. Developing DHS with enhanced electron confinement barriers is critical for improving performance.

Semiconductor grafting, an innovative technology, expands the possibilities for creating heterostructures from dissimilar materials, despite their lattice mismatches and structural differences[18]. This technique utilizes an ultrathin dielectric layer, such as an ultrathin oxide as an interface, which functions both as a double-side passivation and a quantum tunneling layer between two single-crystalline semiconductors[19-26]. This approach facilitates the formation of nearly ideal lattice-mismatched heterojunctions. It has been demonstrated to be effective in applications requiring high power[22, 26], high frequency[27], and advanced optoelectronic capabilities[19, 25], proving its utility and effectiveness.

In this study, we employed the grafting approach to demonstrate a high-quality GeSn-based double heterostructure, enhancing electron confinement and substantially improving the rectification of the GeSn/Ge-based diodes. The Al$_{0.3}$Ga$_{0.7}$As/Ge$_{0.87}$Sn$_{0.13}$/GeSn p-i-n double heterostructure was successfully fabricated using this grafting method. The diode exhibited remarkable performance metrics, including a 1.186 eV conduction band barrier, a mean ideality factor of 1.47, and

Yang Liu and Yiran Li contributed equally to this letter.
Yang Liu, Yiran Li, Jie Zhou, Jiarui Gong, Yi Lu, Daniel Vincent, Samuel Haessly, and Zhenqiang Ma are with the Department of Electrical and Computer Engineering, University of Wisconsin-Madison, Madison, WI 53706, USA. (e-mail: mazq@engr.wisc.edu).
Alireza Abrand and Parsian K. Mohseni are with the Microsystems Engineering, Rochester Institute of Technology, Rochester, NY, 14623. (e-mail: pkmeen@rit.edu)
Sudip Acharya and Shui-Qing Yu are with Department of Electrical Engineering, University of Arkansas, Fayetteville, AR 72701 USA. (e-mail: syu@uark.edu).
This work is supported by Air Force Office of Scientific Research under grant FA9550-19-1-0102 and National Science Foundation under Award No. 2235443.

a mean on/off ratio of 2.95×10³ across 326 devices, featuring a uniformly low density of interface traps ($D_{it}$). Compared to other GeSn heterojunctions developed through CVD epitaxial growth and nanomembrane transfer, our device demonstrated the highest on-off ratio and electron confinement. These comprehensive evaluations affirm the effectiveness of our approach in GeSn-based devices, paving the way for its potential use in electrically pumped GeSn lasers in the near future.

## II. EXPERIMENT

The fabrication process flow of the $Al_{0.3}Ga_{0.7}As$ /$Ge_{0.87}Sn_{0.13}$/GeSn DHS is illustrated in **Fig. 1**. An epitaxially grown AlGaAs stack, consisting of a 100 nm 1×10¹⁹ cm⁻³ p⁺-GaAs on top of the 500 nm 1×10¹⁸ cm⁻³ p⁻-AlGaAs active layer and an $Al_{0.95}Ga_{0.05}As$ scarified layer, was lifted off from its hosting substrate using a polydimethylsiloxane (PDMS) stamp after treatment with 49% hydrofluoric acid (HF), as shown in **Fig. 1(a)(b)**. Meanwhile, the cleaned and polished GeSn epi substrate, containing a 310nm i-$Ge_{0.87}Sn_{0.13}$ layer (post-polishing) and a 600nm GeSn grading buffer layer (8%-13% Sn), was loaded into atomic layer deposition (ALD) system, following a deposition of 5 cycles Al₂O₃ (0.11 nm/cycle) on the surface at 200 °C (**Fig. 1(c)(d)**). The released AlGaAs layer was transfer-printed onto the ALD-coated GeSn substrate via the PDMS stamp (**Fig. 1(e)**), followed by rapid thermal annealing (RTA) at 200 °C for 30 min under N₂ ambient to form chemical bonding (**Fig. 1(f)**). Mesa etching was performed using an inductively coupled plasma (ICP) etcher to open windows on the heterostructure for n-ohmic contacts. Subsequently, n-ohmic contact metal (Ni/Au/Cu/Au: 10/10/100/10 nm) was deposited on the ICP-exposed n⁺-GeSn buffer layer and p-ohmic contact metal (Ti/Pt/Au/Cu/Au: 15/ 50/10/100/50 nm) was deposited on the p⁺-GaAs layer using electron beam metal evaporator (**Fig. 1(g)**). A device isolation dry etching process was then carried out by etching away the surrounding AlGaAs and GeSn following isolation patterning lithography. Finally, all diodes were passivated with 80 cycles of Al₂O₃ (~ 8 nm) using ALD. The schematic structure of the AlGaAs /GeSn/GeSn DHS is shown in **Fig. 1(h)**.

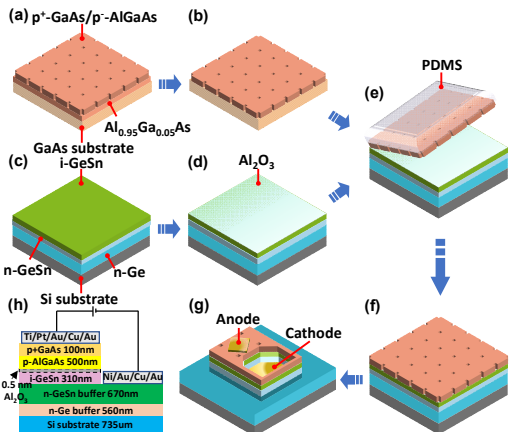

**Fig. 1** Fabrication process flow of the AlGaAs/GeSn/GeSn DHS. (a) p⁺-GaAs/p⁻-AlGaAs stack sample structure after hole etching; (b) $Al_{0.95}Ga_{0.05}As$ scarified layer selectively removed by HF to release the p⁻-AlGaAs layer; (c)GeSn epi structure; (d) 0.5 nm Al₂O₃ deposited on the surface using ALD; (e) Released AlGaAs layer picked up by a PDMS stamp and transfer-printed to the GeSn substrate; (f) Heterostructure formed after RTA; (g) Diode fabricated following conventional photolithography, dry etching, and ohmic contact metal deposition; (h) Schematic illustration of the AlGaAs/GeSn/GeSn DHS.

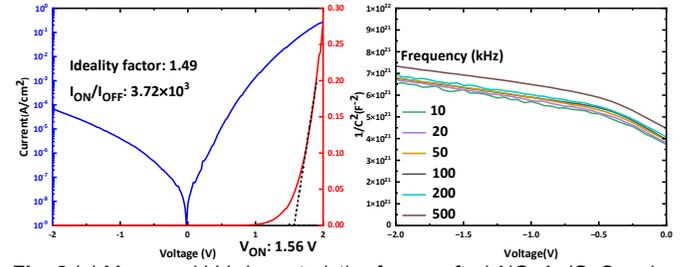

**Fig. 2** (a) Measured I-V characteristics from grafted AlGaAs/GeSn p-i-n DHS in the range of ±2 V (inset: optical image of 4 devices); (b) Measured $1/C^2$ of the diode as a function of bias voltage at multiple frequencies from 20 to 500 kHz.

## III. RESULTS AND DISCUSSION

AlGaAs provides a higher conduction band energy offset than Ge and SiGeSn, along with a lower refractive index compared to GeSn. The detailed band diagram, showing a 1.186 eV conduction band offset as determined by the X-ray photoelectron spectroscopy, can be find in our recent work[ ]. The current-voltage (I-V) characteristics of the AlGaAs/GeSn /GeSn DHS diode at 300K were measured using a Keithley 4200 Semiconductor Parameter Analyzer, with results presented on both linear and semi-logarithmic scales in **Fig. 2(a)**. The inset of **Fig. 2(a)** shows an optical image of 4 devices. To avoid the influence of photocurrent, all measurements were conducted under dark conditions. The diode exhibited significant rectification, with a ratio of 3.72×10³ at ±2 V. At a reverse bias of -2 V, the current density was notably low at 6.9×10⁻⁵ A/cm², indicative of low surface defects and high-quality device fabrication. The I-V characteristics were analyzed using the I-V equation to derive series resistance ($R_s$) [28]. We determined an ideality factor ($n$) of 1.49, a good value largely attributable to the interfacial passivation by Al₂O₃, which effectively reduces $D_{it}$. This low ideality factor confirms excellent carrier injection efficiency within the diode[29, 30]. The series resistance ($R_s$) obtained through analysis was 1.2 kΩ·cm², possibly influenced by the annealing temperature limits of the ohmic contact metal and potential tin segregation at elevated temperatures[31]. Additionally, the turn-on voltage of the diode was derived by fitting the I-V curve in the linear region from 1.5 V to 2 V at a current of approximately 0.15 A, extrapolating to zero to yield a value of 1.56 V. At a forward bias of +2 V, the current density and ON resistance were measured as 0.28 A/cm² and 7.09 Ω·cm².

Furthermore, to understand the $D_{it}$ value between AlGaAs and GeSn , the capacitance-voltage (C-V) characteristics of the $Al_{0.3}Ga_{0.7}As$/$Ge_{0.87}Sn_{0.13}$/GeSn DHS diodes were measured by the Keithley 4200. Notably, as shown in the **Fig. 2(b)**, a negligible frequency dispersion was observed in the C-V curves across a frequency range from 20 kHz to 200 kHz, affirming the device stability[26]. The non-linear behavor of the $1/C^2$ – V plot near 0V indicates the presence of defects in the GeSn layer, likely due to challenges encountered during the growth process[32]. A slight discrepancy observed at 500 kHz likely results from interface defects becoming excited at higher frequencies. Overall, the minimal dispersion in the plot signifies excellent interface quality of the AlGaAs/GeSn and indicates an acceptable $D_{it}$.

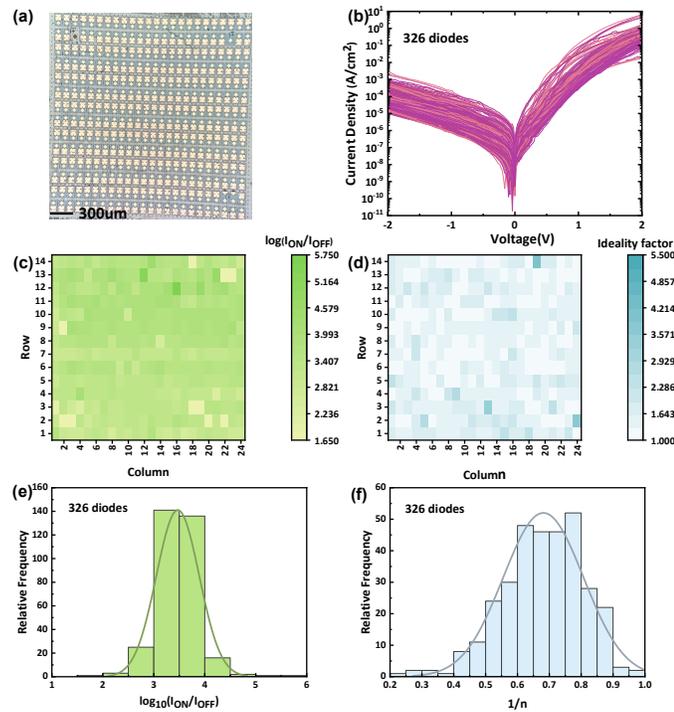

**Fig. 3** (a) Optical micrograph of the fabricated diode array; (b) I-V characteristics across 326 devices; (c)(d) Mapping of the 326 devices rectification ratio and ideality factor; (e)(f) Distribution of the 326 devices' rectification ratio and the reciprocal of ideality factor.

A statistical analysis of the current-voltage characteristics of 326 devices across the entire sample is depicted in **Fig. 3(b)**, measured from the sample shown in **Fig. 3(a)**. The mapping and distribution of the rectification ratios at ±2 V and ideality factors for these 326 diodes are plotted in **Fig. 3(c)(e)** and **Fig. 3(d)(f)**, respectively. Both the base-10 logarithmic rectification ratio ($Log_{10}(I_{ON}/I_{OFF})$) and the reciprocal of the ideality factor ($1/n$) followed Gaussian distributions, with mean values and standard deviations of 3.47 ± 0.42 and 0.68 ± 0.13, respectively. Consequently, the mean diode rectification ratio is calculated to be 2.95×10$^3$ at ±2 V. Using the rule of uncertainty propagation, the mean ideality factor (n) is determined to be 1.47 ± 0.28. The consistency in the diode rectification ratio and ideality factor indicates a uniform interfacial condition, affirming the feasibility of our DHS structure approach for future AlGaAs/GeSn electrically pumped laser device applications. **Fig. 4** benchmarks the conduction band offset and rectification ratio against those from previous studies on similar GeSn p-i-n double heterostructures developed through epitaxy and transfer printing. Our work exhibits a large conduction band offset with the highest on/off ratio[30, 31, 33-37].

## IV. CONCLUSION

In this study, we fabricated and characterized an AlGaAs/GeSn/GeSn double heterostructure diode using the grafting approach. The resulting structure exhibited a 1.186 eV conduction band barrier between AlGaAs and GeSn, a low interfacial density of states, and robust electrical characteristics. This heterostructure ensures adequate electrical confinement, making it a promising approach for fabricating electrically pumped GeSn lasers possibly operating at higher temperatures.

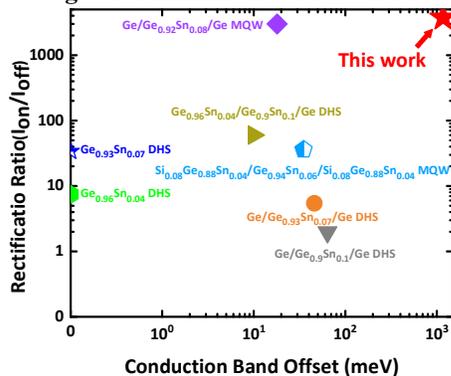

**Fig. 4** Benchmarks of the conduction band offset and rectification ratio for similar GeSn based p-i-n double heterostructures.